# Order/Disorder Dynamics in a Dodecanethiol-Capped Gold Nanoparticles Supracrystal by Small-Angle Ultrafast Electron Diffraction

**Authors:** Giulia Fulvia Mancini[a, 1], Tatiana Latychevskaia[b], Francesco Pennacchio[a], Javier Reguera[c, 2, 3], Francesco Stellacci[c], and Fabrizio Carbone[a, 4*]

**Affiliations**

[a]Laboratory for Ultrafast Microscopy and Electron Scattering, Lausanne Center for Ultrafast Science (LACUS), École Polytechnique Fédérale de Lausanne, CH-1015 Lausanne, Switzerland.

[b]Physics Institute, University of Zurich, Winterthurerstrasse 190, 8057 Zurich, Switzerland.

[c]Supramolecular Nanomaterials and Interfaces Laboratory, Institute of Materials, École Polytechnique Fédérale de Lausanne, CH-1015 Lausanne, Switzerland.

[1] Present address: JILA, University of Colorado, Boulder, CO 80309-0440.

[2] Present address: CIC biomaGUNE, Paseo de Miramón 182C, 20009 Donostia-San Sebastian, Spain.

[3] Present address: Ikerbasque, Basque Foundation for Science, 48011 Bilbao, Spain.

[4] To whom correspondence should be addressed: Prof. Fabrizio Carbone, Laboratory for Ultrafast Microscopy and Electron Scattering (LUMES). Faculty of Basic Sciences (SB), École Polytechnique Fédérale de Lausanne (EPFL), EPFL campus, CH H2 595, Station 6, CH-1015 Lausanne. Switzerland.

Email: fabrizio.carbone@epfl.ch

Phone: +41 21 69 30562

**Authors contributions:** F. C. and F.S. designed research; G.F.M. and F.P. performed research; G.F.M. and T.L. analyzed data; G.F.M., T.L. and F.C. wrote the paper; J.R. prepared the samples.

**Abstract**

The design and the characterization of functionalized gold nanoparticles supracrystals require atomically-resolved information on both the metallic core and the external organic ligand shell. At present, there is no known approach to characterize simultaneously the static local order of the ligands and of the nanoparticles, nor their dynamical evolution. In this work, we apply femtosecond small-angle electron diffraction combined with angular cross-correlation analysis, to retrieve the local arrangement from nanometer to inter-atomic scales in glassy aggregates. With this technique we study a two-dimensional distribution of functionalized gold nanoparticles deposited on amorphous carbon. We show that the dodecanethiol ligand chains, coating the gold cores, order in a preferential orientation on the nanoparticle surface and throughout the supracrystal. Furthermore, we retrieve the dynamics of the supracrystal upon excitation with light, and show that the positional disorder is induced by light pulses, while its overall homogeneity is surprisingly found to transiently increase. This new technique will enable the systematic investigation of the static and dynamical structural properties of nano-assembled materials containing light elements, relevant for several applications including signal processing and biology.

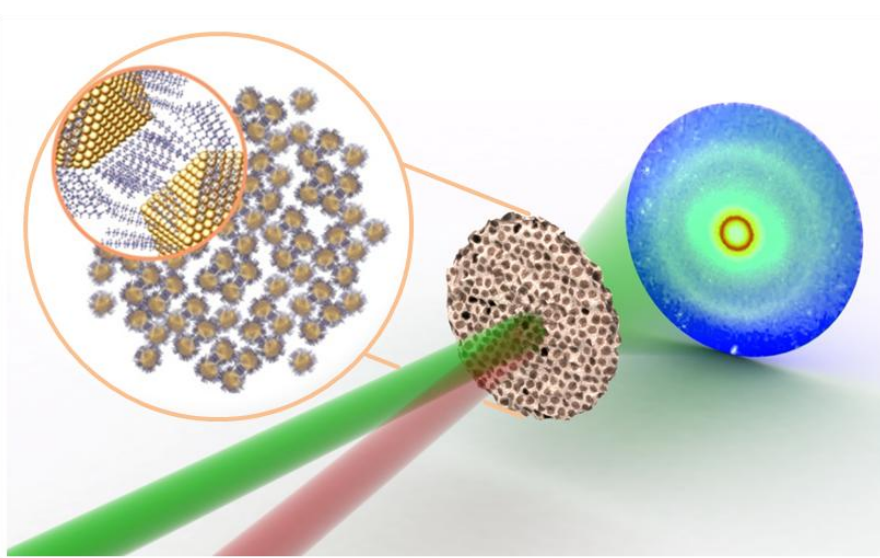

**Table of Contents** graphic to the manuscript

1. **Introduction**

In metal nanoparticles (NPs)[1-3], the metallic core provides some key properties, *e.g.* magnetization, plasmonic response or conductivity, with the ligand molecules giving rise to others like solubility, assembly or interaction with biomolecules[4, 5]. The formation of NPs supracrystals depends on a complex interplay between many forces, some stemming from the core, some from the ligands[6]. These assemblies are promising candidates for many applications in very different fields, such as electronics and medicine[7, 8].

The structural dynamics of non-functionalized polycrystalline gold thin films has been investigated by means of Ultrafast Electron Diffraction (UED) in transmission[9, 10] and time-resolved X-ray diffraction (XRD)[11, 12]. Direct structural information on the melting and crystallization dynamics of gold NPs suspended on organic membranes was obtained by UED[13], while coherent acoustic phonons have been observed in single bare gold NPs by means of time-resolved three-dimensional imaging in a X-ray Free-Electron Laser (X-FEL)[14]. All these experiments were limited at the observation of the intra-NP dynamics, and none dealt with functionalized NPs.

While electrons offer a higher cross-section for interaction with matter than X-rays, yielding a good sensitivity in thin samples containing light elements[15, 16], their limited transverse coherence complicates the observation of large distances in diffraction[17]. Conversely, X-ray beams have the required coherence, but smaller interaction cross-section, making it difficult to observe the dynamics of the chains of carbon and hydrogen atoms constituting the ligands.

Structure retrieval methods using the angular Cross-Correlation Function (CCF) analysis of an XRD pattern have demonstrated the possibility to retrieve information on the sample local order and symmetry[18, 19]. Latychevskaia *et al.*[20] reported that for a dense system of identical particles, even at a very low coherence length of the probing wave (comparable to the size of the single particle), the characteristic modulations of certain symmetries in the sample appear in the CCF.

Small-angle femtosecond (fs)-electron diffraction experiments presented in this work can access the length-scale between few Ångstrom (Å) to a few nanometers (nm) with fs time-resolution and sensitivity to the light elements in ligands. To do this, a trade-off between the beam brightness and its coherence at the sample was found in our high-flux ($10^9$ e⁻/sec) UED set-up[21]. The CCF analysis has been applied and revealed the ordered arrangement of the ligands binding to the NPs in the two-dimensional supracrystal, even though the NPs themselves are not arranged in a perfect lattice. The discovered arrangement of the ligands indicates the presence of a specific preferred orientation of the NPs in the two-dimensional supracrystal. Furthermore, time-resolved experiments clarify the way in which light-induced thermal disorder evolves in such systems, revealing the time-scales involved in the entropy variations of the distribution of the NPs in the supracrystal and of the ligands sublattices.

## 2. Set-up and experiment

1-dodecanethiol-coated NPs were synthesized using a modification[22] of the method described by Zheng *et al.*[23]. A NPs monolayer was prepared by a drop-wise deposition of a toluene solution of NPs onto the water subphase of a Langmuir trough. After 10 min the layer was compressed at 10mm/min until it reached 18mN/m (at the isotherm solid phase region) and then it was transferred to an amorphous carbon-coated grid through a Langmuir-Schaefer deposition. A Transmission Electron Microscopy (TEM) image of the sample, characterized by an average NPs core diameter of 5.7nm and a polydispersity of 9%, is shown in Fig. 1a, where crystallographic-like planes in the NPs arrangement are indicated. In this TEM image, the ligands are not visible as they do not provide enough contrast for 200keV electrons. Previous experiments demonstrated that the ligands may be observed at lower voltages[24, 25] for isolated particles. Recently, the ligands orientation on an ensemble of oleic-acid-coated PbSe nanoparticles has been imaged in a HAADF-STEM operated at 200kV[26]. However, none of these experiments has provided, at once, static and dynamic information on ligands, as they were limited to the observation of the system in stationary conditions.

A KMLabs Wyvern Ti:sapphire amplified laser generating 50-fs (FWHM), 700-µJ pulses, 800nm central wavelength, 20kHz repetition-rate, was employed to generate both the probing electrons (30keV, $\lambda$=7pm) and the photo-excitation. The temporal spread of the 30kV probe electron pulses was controlled by means of a radiofrequency (RF) compression cavity[27, 28], allowing to store up to $6 \cdot 10^5$ electrons in ~300fs/160µm bunches at the sample[21]. A sketch of our set-up is presented in Fig. 1b. Fs snapshots of the diffraction pattern from the NPs supracrystal were recorded at different time-delays between the pump photo-excitation and the probe, building a fs-resolved movie of the dynamics of the sample. Experiments were performed in transmission geometry at room temperature and with an almost collinear arrangement between the pump and probe pulses, in order to reduce the spatio-temporal mismatch between them. The background pressure in the experimental vacuum chamber was below $10^{-9}$ mbar. Photoinduced changes in the sample were initiated by 800nm (1.5eV) pump-pulses focused to a spot of 220µm, which were temporally overlapped on the sample to a probing electron beam with 160 µm-diameter. The incident fluence on the sample was 10mJ/cm$^2$, while the effective fluence absorbed by the sample is estimated around 100µJ/cm$^2$. This latter takes into account for the optical reflectivity of gold in a layer of 7nm thickness, for the penetration depth of gold for electrons, estimated around 7-8nm at 1.5eV[29], and for the sample density.

The diffraction patterns were acquired with a very small current (320.4pA for every time delay, 4.11µC as total charge on the NPs with this amount being distributed over several electron pulses impinging on the sample) thanks to the high sensitivity camera and the pulsed electron beam. Also, the very low duty cycle allows a large relaxation time between subsequent pulses. For these reasons, radiation damage was not observed in these experiments (see Supporting Information). The diffraction pattern, shown in Fig. 1c, was formed on a phosphor screen and was recorded by a charge-coupled device camera capable of single electron detection; in the experiment, the sample-to-camera distance was optimized so that the transverse coherence was comparable to the distances of interest in the sample (~5nm), corresponding to small deflection angles in the *mrad* range.

## 3. Static characterization of the supracrystal

The investigation of the local symmetry in our sample is carried out at the three different scattering vectors marked in the diffraction pattern of Fig. 1c, which relate to three different length-scales in the sample. The low order feature $s_1$ is attributed to the arrangement of the NPs in the supracrystal, where the NPs form crystallographic planes with a distance $d_1$=6.6nm as depicted in the upper inset of Fig. 1a and sketched in the rendering of Fig. 2a.

The features at higher scattering vectors, labeled $s_3$ and $s_4$, correspond to the arrangement of the ligands binding to the surface of each gold NPs core. The reciprocal distance $s_3$ refers to the hexagonal superlattice of dodecanethiol ligands chains that are binding the gold core facets at specific locations between gold atoms, as shown in the inset of Fig. 2b. In the gold face center cubic (*fcc*) crystal structure, the distance between two neighboring gold atoms is 2.88Å, which leads to a distance between two nearest locations of the ligand attachment of 4.99Å (see inset of Fig. 2d). This gives the distance between the crystallographic planes created by the carbon chains of $d_3 = \frac{\sqrt{3}}{2} \cdot 4.99$ Å = 4.33Å. A typical geometrical arrangement of the system in static conditions is the one in which the ligands attached to one facet are stretched and the NPs are randomly oriented throughout the sample, as explained in details in the Supporting Information (SI). In this situation, a diffraction ring at $s_3$=1.45Å$^{-1}$ is expected in the diffraction pattern.

The signal at the scattering vector $s_4$=1.68Å$^{-1}$ is related to the real space distance $d_4$=3.72Å, which is the distance between chains of carbon atoms in ordered ligands when those are projected from a specific direction of view, as depicted in Fig. 2g. This orientation of ligands relates to a peculiar orientation on the NPs which turns out to be the preferred orientation of NPs throughout the sample, as discussed later in the text.

### 3.1. Cross-Correlation function and Fourier analysis of the diffraction pattern

Local symmetries in the sample can be revealed by analyzing the modulations in the CCFs at the related scattering vectors $s_j$ (*j*=1, 3, 4). The normalized CCF is defined as[18, 19]:

$$C_{\text{norm}}(\Delta) = \frac{\langle I(s,\varphi)I(s,\varphi+\Delta)\rangle_\varphi - \langle I(s,\varphi)\rangle_\varphi^2}{\langle I(s,\varphi)\rangle_\varphi^2} \qquad (1)$$

where $I(s,\varphi)$ represents the scattered intensity at the defined scattering vector *s* and the angle $\varphi$; $\Delta$ is the shift between the two angles, as indicated in Fig. 1c and $\langle \ \rangle_\varphi$ denotes an averaging over $\varphi$. The details on the formalism of the angular cross-correlation analysis are reported elsewhere[30, 31]. The non-normalized CCF is calculated as:

$$C(\Delta) = \langle I(s,\varphi)I(s,\varphi+\Delta)\rangle_\varphi = \text{Re}\left(F_\varphi^{-1}\left(\left|F_\varphi\{I(s,\varphi)\}\right|^2\right)\right), \qquad (2)$$

where $F_\varphi\{...\}$ is the one-dimensional Fourier Transform (FT) over the $\varphi$ coordinate. The Fourier transform of $I(s,\varphi)$ yields a spectrum where the most pronounced frequencies indicate the presence of a signal with a corresponding periodicity. In our data, the low-order frequency $\nu=2$ is assigned to the beam astigmatism and to the signal from the amorphous carbon membrane supporting the sample, and it is filtered in the analysis of the CCF (SI).

## 3.2. Experimental diffraction pattern

In Fig. 2a – c, the results of the CCF analysis at the scattering vectors $s_1$ are reported. Such a distance in the reciprocal space corresponds to the planes of the supracrystal, rendered in Fig. 2a. The spectrum of the intensity profile at $s_1$ (inset in panel b) is dominated by the frequency $\nu=2$ followed by $\nu=6$ and $\nu=4$, as shown in Fig. 2b. The $\nu=2$ harmonic originates from the inherent slight astigmatism of the electron beam and the signal from the amorphous carbon of the substrate, as discussed in details in the SI. After filtering out $\nu=2$, a four-fold modulation of the CCF is observed, see Fig. 2c (red curve). Next, after filtering out $\nu=2$ and $\nu=4$, a six-fold periodicity is retrieved in the CCF, as visible in Fig. 2c (purple curve). At the reciprocal space distance $s_3$ corresponding to the arrangement of ligands in a superlattice on the NPs facets, rendered in Fig. 2d, the CCF yields a clear four-fold modulation after filtering out the $\nu=2$ Fourier component, as shown in Fig. 2f (red curve). The absence of a six-fold arrangement at this distance is evident from the CCF (purple curve) shown in same figure. At $s_4$, corresponding to the distance between the carbon chains in the ligands, rendered in Fig. 2g, the CCF shows a clear four-fold modulation. When both the $\nu=2$ and $\nu=4$ frequencies are filtered out, a six-fold CCF profile (Fig. 2i, purple solid line) is observed.

3.2.a Background analysis.

Among the retrieved periodicities, we determined which modulations originate solely from the scattering of electrons from the gold NPs and their ligands, and which ones stem from the substrate. It is essential to verify that the diffraction intensity modulations in the experimental diffraction pattern are not coming from the TEM grid supporting the supracrystal or the substrate, but are a signature of the sample itself. Therefore we applied the CCF and Fourier transform analysis to the experimental images acquired only from the substrate, without the sample. We compared the results from the dodecanethiol-coated NPs supracrystal with the ones obtained following the same acquisition process and data analysis on an amorphous carbon coated TEM copper grid and on an empty copper TEM grid. The static data from the supporting

TEM grid in terms of electron flux, exposure time and experimental geometry are identical to the ones reported for the acquisition of the electron diffraction pattern from the supracrystal, while those from the empty grid are obtained with much lower exposure time because the lower transmissivity of such a sample would cause the CCD saturation. The results, reported in Fig. 3 and Fig. 4, show that:

(i) The six-fold modulations of the intensity retrieved at $s_1$ and $s_4$ (Fig. 2c and Fig. 2i respectively, purple curves) and assigned to the local order within the supracrystal structure, are not found at $s_1$ and $s_4$ in the intensity of the diffraction pattern of the substrate. Thus, the six-fold symmetries within the sample retrieved from the modulations in the CCFs at $s_1$ and $s_4$ can be safely attributed to the NPs supracrystal.

(ii) By comparison of the CCF at $s_1$ shown in Fig. 2c (red curve) with that of the amorphous carbon shown in Fig. 3b (red curve), it turns out that both the supracrystal and the amorphous carbon show a similar four-fold CCFs, that are also comparable in amplitude. Thus, we can attribute this four-fold symmetry in the CCFs to the tetragonal arrangement of the amorphous carbon in the substrate.

(iii) The four-fold symmetry retrieved at $s_4$ (Fig. 2i, red trace) from the diffraction pattern of the sample is not found at the same $s_4$ in the diffraction patterns, either of the amorphous carbon-coated or of the empty grid. Thus, this four-fold modulation at $s_4$ solely originates from the NPs local arrangement in the supracrystal.

(iv) The four-fold modulation retrieved at $s_3$ in the data from the amorphous carbon-coated grid (Fig. 3d, red trace) confirms that the substrate contributes to a four-fold modulation at $s_3$ (Fig. 2f, red trace).

(v) Both the NP supracrystal and the amorphous carbon samples show a strong and distinct two-fold modulation at $s_1$, while the diffraction pattern of the empty TEM grid was overexposed at those $s$. The slight astigmatism of the beam is known from the original characterization of our experimental setup[21] and would also contribute to a two-fold modulation in the CCF.

The results of this analysis are summarized in Table 1, where we show that the modulations of the CCF relative to the NPs arrangement in the supracrystal ($s_1$) and the distance between chains of carbon atoms in ordered ligands ($s_4$) can be solely attributed to the sample properties and not to either the amorphous carbon substrate (Fig. 3) or the copper support grid itself (Fig. 4).

## 3.3 Simulated diffraction patterns. Sphere Lattice Model and Superposition of Atomic Scattering Amplitude simulations.

To assign and explain the symmetries observed in the diffraction pattern to the corresponding real space objects, a series of simulations was carried out. Figure 5a displays the diffraction pattern simulated as the squared amplitude of the FT of the TEM image of the sample (Fig. 1a), covering the scattering vector range up to a value of $0.4\text{Å}^{-1}$. To mimic the distribution of the NPs in the supracrystal we used the Sphere Lattice Model (SLM) simulation, where the NPs are represented by a sphere[32], and their arrangement is governed by two parameters: NP diameter and deviation from the position in a perfect hexagonal lattice.

Two sample distributions simulated by the SLM at different disordering conditions are shown in Fig. 5b and c, with the corresponding Fourier transforms (FTs). The radial average intensity has been calculated as follows:

$$I(s) = \frac{1}{2\pi} \int I(s,\varphi)\, d\varphi. \tag{3}$$

The radial intensity profiles for the two simulated and the experimental TEM-derived diffraction patterns are displayed in Fig. 5d. The parameters of the SLM model were varied until the best match between the position of the peaks in the simulation and the experimental curve was achieved. The best match was found to be for a sphere diameter of 5.7nm, and a hexagonal arrangement with a core-to-core distance of 7.6nm. These parameters give the distance $d_1$=6.6nm between the planes of crystallographically arranged NPs in the supracrystal (upper inset of Fig. 1a). The real space distances, $d_1$ and $d_2$, are introduced in the upper and lower panels of Fig. 1a, respectively.

In Fig. 5e, we show the CCF curves of the intensity distribution at a scattering vector $s$ corresponding to the first order of diffraction from the planes with spacing $d_1$ ($s_{1,1}$). The three CCFs show a six-fold modulation, confirming the hexagonal arrangement of the NPs in the supracrystal. In the experimental electron diffraction pattern (Fig. 1c, to be distinguished from the FT of the TEM image), the diffraction at $s_{1,1}$ is overexposed by the central beam. However, the higher diffraction orders of $d_1$ can be detected at higher $s$ values. For this reason, the analysis at $s_1$ ($d_1$=6.6nm) in the experimental electron diffraction pattern was carried out at the seventh diffraction order $s_{1,7}$ ($s_{1,7}=n\cdot 2\pi/d_1=$ 7·2$\pi$/6.6nm= $0.66\text{Å}^{-1}$), as marked in Fig. 1c. Further discussion on this point is provided in the SI. The six-fold CCF obtained for the experimentally measured intensity at $s_{1,7}$ (Fig. 2c, purple curve) agrees well with the six-fold CCFs obtained from simulated diffraction patterns (Fig. 5e) for the lower diffraction orders of $s_1$. Thus, different diffraction orders from the same real-space object show an identical CCF periodicity. The simulations, the FT of the TEM image and the small angle diffraction data, all confirm that the NPs are arranged into a hexagonal lattice. Furthermore, the comparison of the SLM simulations to the experimental data allows us to quantitatively estimate the disorder in the crystal. The four-fold modulation in the CCF retrieved from experimental intensity at $s_1$, and displayed in Fig. 2c (red curve), is observed both in the sample with gold NPs and in an empty amorphous carbon coated grid. Therefore, the interpretation of this four-fold symmetry cannot be done with certainty as such a four-fold modulation could originate either from the amorphous carbon or from an inhomogenity in the supracrystal having tetragonally distorted domains.

The shorter distances, $d_3$ and $d_4$, are related to the atomic arrangement of gold cores and ligands. The simulated diffraction pattern at these distances is obtained by the Superposition of Atomic Scattering Amplitudes (SASA) from individual atoms in the NPs, where the shape of the NPs and the orientation of the ligands on their surface are modelled according to previous reports[32-34]. It is worth noting that only an ordered structure gives rise to a characteristic peak in the diffraction pattern. For example, the wave scattered by *gauche* ligands does not create a characteristic peak but contributes to the background intensity. On the other hand, a bunch of equally oriented *trans* ligands constitutes a lattice of rods, which can give rise to characteristic scattering peaks in the diffraction pattern. For this reason, a single bunch of *trans* ligands attached to a gold facet was isolated for simulation, as depicted in Fig. 5f. To simulate all the possible orientations of the bunch of ligands, and thus all possible orientations of the gold facets, the bunch was randomly rotated around its centre. At each bunch rotation, for each atom, its $x$, $y$, $z$ coordinates were assigned and the complex-valued scattering amplitudes specific to each chemical element (Au, S, C, H) were calculated using the NIST library[35] for 30keV electrons. The results of the SASA simulation are reported in Fig. 5f–i. The complex-valued waves scattered off each atom were superimposed in the far-field and the intensity of the total wave field provided the diffraction pattern. For each rotation of the bunch such a simulated diffraction pattern was created. One hundred random rotations were calculated and the related simulated diffraction patterns were added together (incoherent addition). Because of the stronger contribution from the carbon atoms, and of their intense scattering amplitude in the forward direction, their signal dominates the resulting diffraction pattern.

Figure 5f shows a magnified region of the radial intensity distribution calculated with equation (3). The $I(s)$ obtained from the simulated diffraction pattern of randomly oriented bunches has a maximum at $s_3=1.45$Å$^{-1}$ (Fig. 5f, blue curve). The CCF analysis of the intensity at $s_3$ in the simulated diffraction pattern (not shown here) does not reveal any modulations, which is predictable, as the bunch of ordered *trans* ligands was randomly rotated. On the other hand, a four-fold modulation is revealed at $s_3$ in the experimental diffraction pattern (Fig. 2f), which can be attributed to tetragonal arrangement of the amorphous carbon in the substrate.

Next, in the simulation, the bunch of *trans* ligands was controllably rotated and a specific orientation of ligands with respect to the electron beam was achieved, which gives the diffraction peak at $s_4$. This arrangement is depicted in Fig. 5f and in Fig. 5g and named "preferred orientation"; the electron beam hits the sample orthogonally to the figure plane. "Preferred orientation" refers to the NP orientation under a specific polar angle, whereas the orientation of the NPs in the sample plane is arbitrary. A single bunch of ligands creates two peaks in the diffraction pattern, as shown in Fig. 5g. When the NPs are arranged in a hexagonal lattice, as shown in the corner of Fig. 5h, where each NP is in "preferred orientation", there are six possible orientations of the NPs in the sample plane, resulting in three possible orientations of bunch of ligands and hence six peaks in the diffraction pattern, see Fig. 5h. When the NPs are arranged in a tetragonal lattice, as shown in the corner of Fig. 5i, with each NP in "preferred orientation", there are two possible orthogonal orientations of bunch of ligands and hence four peaks in the diffraction pattern, see Fig. 5i. Thus, this analysis demonstrates the presence of regions with tetragonal NPs arrangement within regions with hexagonal NPs arrangement in the supracrystal. The cross-validation between the simulations and the experimental data shows a good agreement: a six-fold modulation was discovered at $s_4$ in the experimental diffraction pattern (Fig. 2i), and no six-fold modulations were found at $s_3$ (Fig. 2f).

### 4. Dynamics

Following the static characterization of the system, we studied the behaviour of the NPs supracrystal under photoexcitation. In particular, we selected $s_1$ ($s_{1,7}$) to study the re-arrangement of the NPs within the supracrystal, and we select $s_4$ to study the local re-arrangement of the ligands. The radial intensity curves at the selected scattering vector $s$ have been retrieved following equation (3) and their amplitude as function of the time delay is shown in Fig. 6a.

The intensity traces as a function of time were normalized to their average value at negative times ($t<t_0$) and fitted to a mono-exponential curve. The energy deposited by the pump on the sample causes thermal disorder in the supracrystal, evidenced by a decrease of intensity at $s_1$ after the photo-excitation (blue curve, diamonds), with a time scale of 12±1ps. As a result, the supracrystal local hexagonal lattice arrangement is perturbed. The intensity at $s_4$ also decreases after the photoexcitation (red curve, circles), as a consequence of photo-induced thermal disorder (10±1ps). However, while the intensity of peak at $s_1$ drops by 10% the intensity of peak at $s_4$ drops by only 2%, which means that although the NPs are slightly rearranging themselves, the ligands on their surfaces preserve the preferential arrangement shown before the photo-excitation. The recovery time for the re-arranged system is beyond the probed time-scale range, *i.e.* longer than 220ps after light exposure.

The evidenced dynamics is also confirmed by studying the CCF at both $s_1$ and $s_4$ at different time delays ($t_1$, $t_2$, $t_3$), as shown in Fig. 6b-d, and Fig. 6e-g. Here the CCFs were calculated as:

$$C(\Delta) = \langle I(s,\varphi)I(s,\varphi+\Delta)\rangle_\varphi - \langle I(s,\varphi)\rangle_\varphi^2 \tag{4}$$

where the normalization factor $\langle I(s,\varphi)\rangle_\varphi^2$ at the denominator was eliminated from the calculation in order to retrieve the absolute amplitude of the CCFs and to study its time dependence. The six-fold modulations in the CCF for both $s_1$ and $s_4$ are reported at three selected time delays: before the photo-excitation ($t_1$, Fig. 6b and e), right after the photo-excitation ($t_2$, Fig. 6c and f)), and several ps after photo-excitation ($t_3$, Fig. 6d and g). The amplitude decrease of the CCFs at both scattering vectors as a function of time is consistent with the dynamics in the system retrieved from the time traces of Fig. 6a.

Remarkably, the six-fold modulation of the CCF at both $s_1$ and $s_4$ is found to transiently become more evident as higher signal-to-noise ratio (SNR) upon photoexcitation, despite decreasing in amplitude. As demonstrated by Latychevskaia *et al.*[20], the amplitude of the CCF is proportional to the number of the domains, where the scatterers are organized into an ordered lattice. The experimental traces reported in Fig. 6b-d ($s_{1,7}$) and Fig.6e-g ($s_4$) show that the CCF amplitude is decreasing with time, indicating that the number of locally ordered regions is decreasing. However, the appearance of the peaks in the CCF is given by ordering in the orientation of the domains. A sample with equally oriented domains results in CCF peaks with high SNR, while a sample with more randomly oriented domains results in more noisy CCF peaks. This is demonstrated with the following SLM simulations. The perfect lattice arrangement of the NPs, Fig. 5b, is modified by selecting round domains and rotating them by an angle η which is Gaussian distributed with the mean=0° and standard deviation σ=20° (Fig. 6h) and σ=4° (Fig. 6i). The domains are selected to have a size of 60nm and a centre-to-centre distance between domains of 80nm. For each simulation the corresponding FT is reported in Fig. 6h and i. The CCFs retrieved at $s_{1,1}$ for both FTs and for the perfect lattice arrangement are compared in Fig. 6j-l. Upon increasing the degree of order in the supracrystal, by decreasing the rotation angle of each domain, peaks in the diffraction pattern become sharper. Consistently, the six-fold CCF, which shows broad peaks characterized by the presence of satellites for the [20°, 0] case (Fig. 6j), becomes sharper and more pronounced in the [4°, 0] case (Fig. 6k), all the way to the perfectly ordered lattice (shown in Fig. 5b), in which the CCF consists of six sharp peaks (Fig. 6l). In our experiment, the SNR of the peaks in the CCF distribution for both $s_1$ (Fig. 6b-d) and $s_4$ (Fig.6e-g) reaches its maximum at $t_2$, after the photo-excitation, showing that the local re-arrangement is triggered simultaneously for both the supracrystal and the ligands. The transient increase in the SNR of the CCF evidences that the system reaches some state with the maximal ordering of the domains but then it relaxes to a more favorable state, where some of NPs and the ligands are not ordered. Thus, while the photo-excitation process induces disordering within the supracrystal, the homogeneity of the distribution of the scattering improves, which in turn means that light excitation causes an overall disorder but also induces annealing of the grains in the sample.

## 5. Conclusions

In disordered elastic media, elastic forces of different origin, *e.g.* chemical, magnetic or electrostatic, provide the binding necessary to create order[36]. Their interplay with disorder, of thermal or quantum origin, gives rise to universal behaviours whose investigation has been mostly limited to the observation of frozen systems in static conditions[37, 38]. However, ordering and disordering phenomena are essentially dynamical. In our study, we demonstrate that it is now possible to determine the interplay between disorder and elastic forces with a spatial resolution that allows distinguishing every microscopic constituent of the system and a temporal resolution in the time scale of their motions.

**Acknowledgements**

The experimental work was funded by the Swiss National Science Foundation (SNSF) through the grant No. PP00P2–128269/1. The authors acknowledge B. Patterson and A. Al Haddad for useful discussions. The authors declare no competing financial interest.
**Associated content:**

Supporting Information Available: Spheres Lattice Model (SLM) simulations; Superposition of Atomic Scattering Amplitude (SASA) simulations and arrangement of ligands on the NP surface; Crystallographic arrangement of NPs in the supracrystal: diffraction and angular cross-correlation functions at $s_1$ for higher diffraction orders; Beam astigmatism and stability of the properties of the electron beam; Analysis of the Fourier components as function of $s$-vector coordinate; Cross- validation of the absence of sample damaging (PDF). This material is available free of charge via the Internet at http://pubs.acs.org.

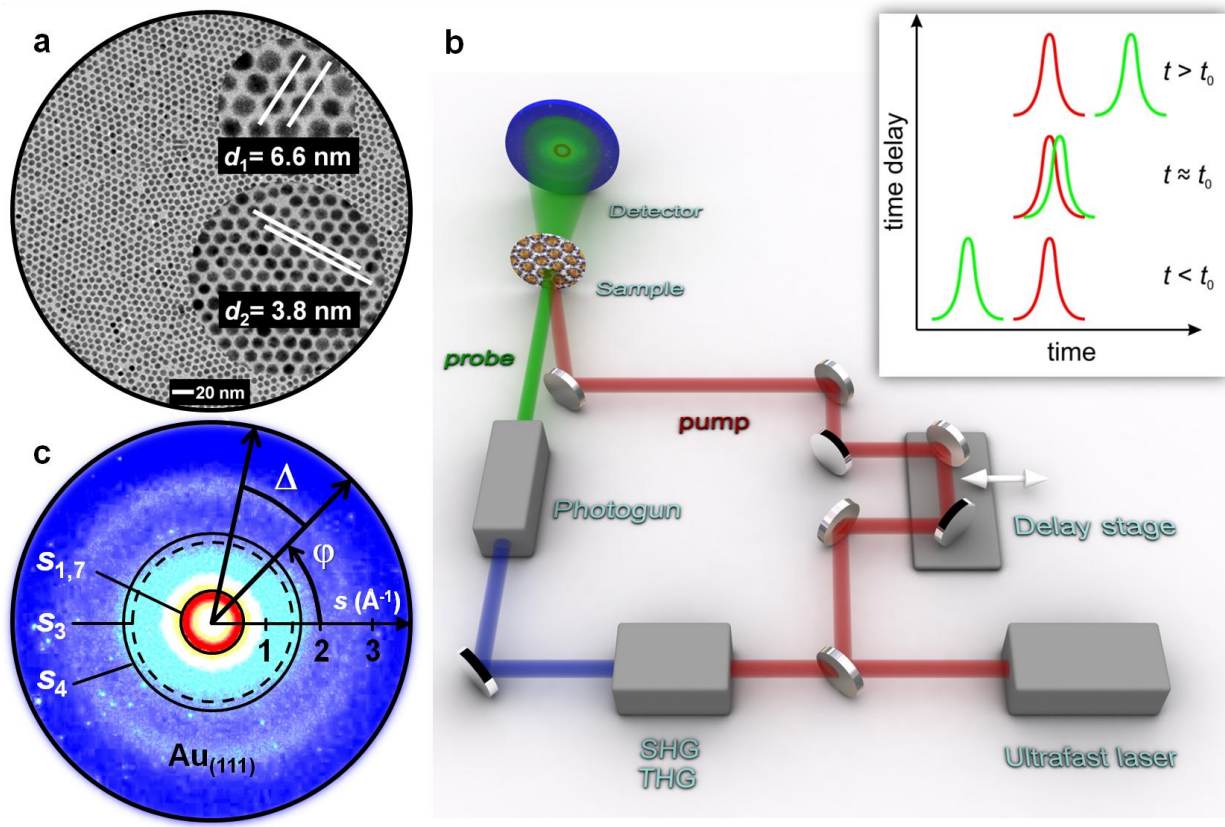

**Figure 1.** UED setup and principle for small-angle time-resolved electron diffraction. (a) TEM image of the sample in static conditions. Insets: Arrangement of the NPs and crystallographic planes with distances $d_1$ and $d_2$. (b) Schematic layout of the UED experiment. A 160μm-diameter electron beam with partial coherence of 5nm and 7pm wavelength is impinging on the sample which is at a distance of 23cm from a charge coupled device camera with 47μm pixel size. For each time-delay 740 diffraction patterns containing $2 \cdot 10^5$ electrons each have been acquired and summed. Inset: The ultrashort light and electron pulses are arranged in a properly timed sequence with the use of a delay stage, and the sequence of pulses is repeated timing the electron pulse to arrive before or after the laser pulse. In this way snapshots of the electrons diffracted from the sample are recorded in a stroboscopic fashion. (c) Experimental electron diffraction pattern of the sample at $t < t_0$. The investigation of the local symmetry in our sample is carried out at the three marked scattering vectors $s_{1,7}$ (referred to as $s_1$ within the text), $s_3$ and $s_4$, as discussed in the text. The azimuthal angle $\varphi$ and the angle shift $\Delta$, employed in the calculation of the CCF, are also reported in the diffraction pattern.

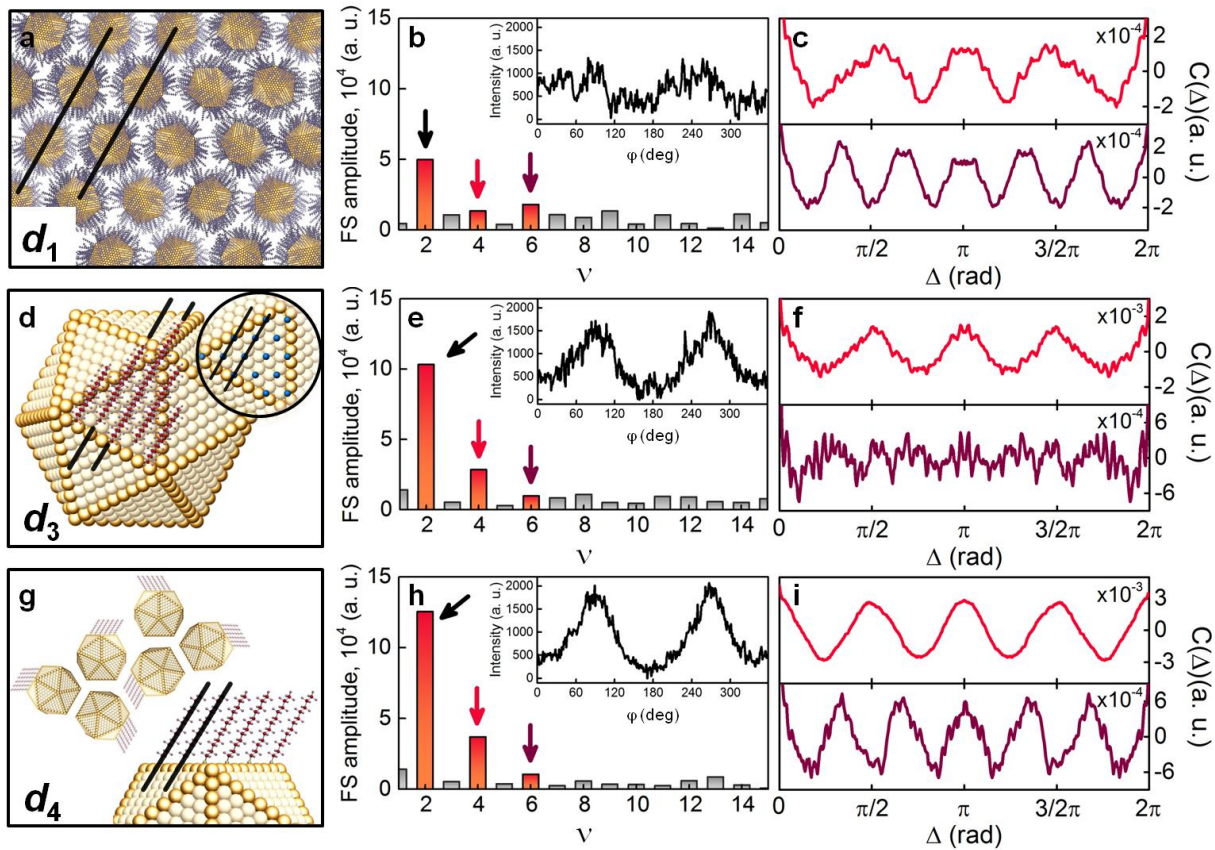

**Figure 2.** Static Fourier spectrum analysis at $d_1$, $d_3$ and $d_4$. (a, d, g) Rendering of the real space objects corresponding to the distances $d_1(s_1)$, $d_3(s_3)$ and $d_4(s_4)$, respectively. (b, e, h) Fourier spectrum obtained by the Fourier transform of the angular intensity profile (insets, black curves). (c, f, i) Upper, red curves: CCF obtained by setting to zero the frequency ν=2 in the Fourier spectrum. Lower, purple curves: CCF obtained by setting to zero the frequencies ν=2 and ν=4 in the Fourier spectrum. All the Fourier spectra as well as the angular intensity curves show a ν=2 frequency. This behaviour is assigned to the astigmatism of the electron beam and the presence of amorphous carbon in the sample support, which is the subject of extended discussion in the SI. The CCF curves have been obtained from the raw data following dark noise subtraction and following the procedure explained in the text. The "large" values of the CCF are motivated by the presence of local ordering throughout the entire sample, as theoretically demonstrated in (20).

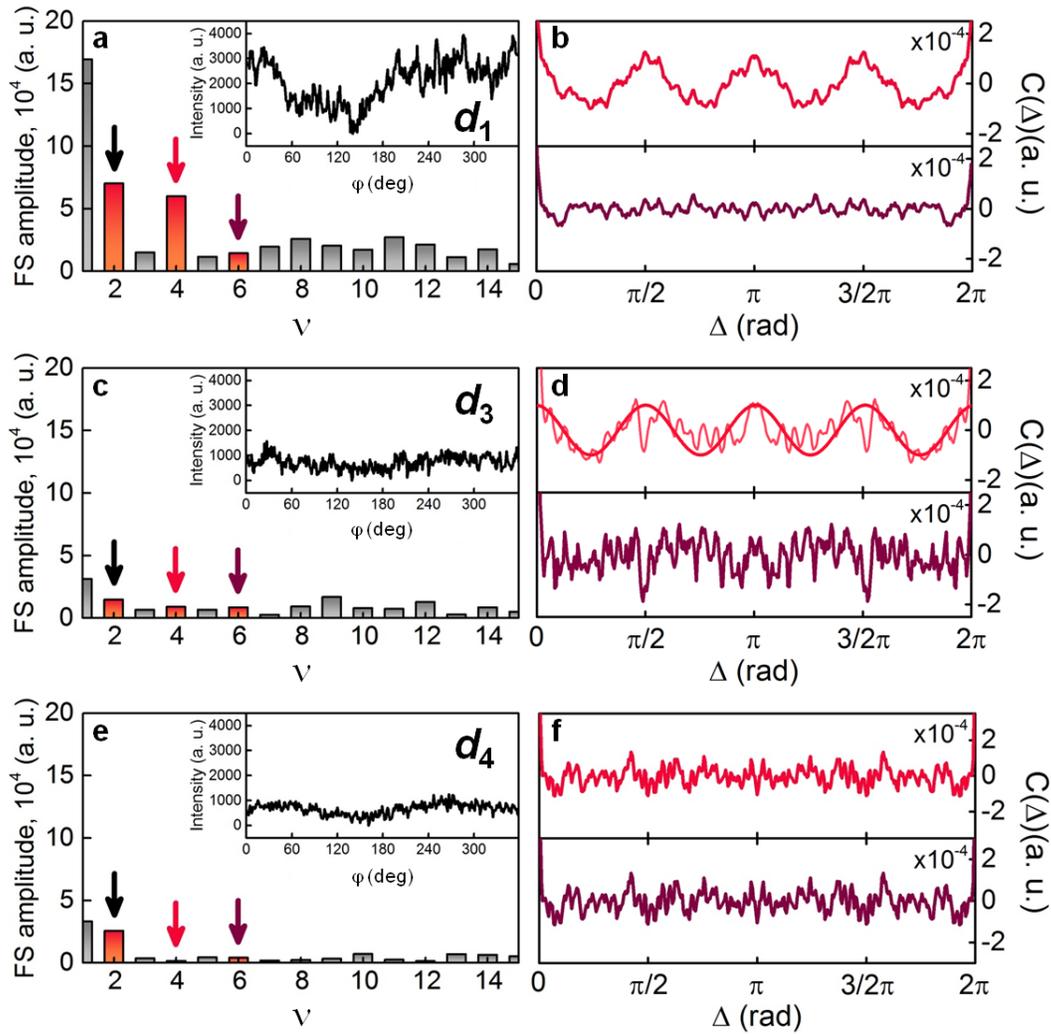

**Figure 3.** Fourier spectra and CCFs at $s_1$, $s_3$ and $s_4$ in the amorphous-carbon coated TEM grid. (a, c, e) Fourier spectra obtained by the Fourier transform of the angular intensity profile (insets, black curves). (b, d, f) Upper, red curves: CCF obtained by setting to zero the frequency ν=2 in the Fourier spectra. Lower, purple curves: CCF obtained by setting to zero the frequencies ν=2 and ν=4 in the Fourier spectra. The data show clearly that the information retrieved from the supracrystal are different from the ones of the substrate.

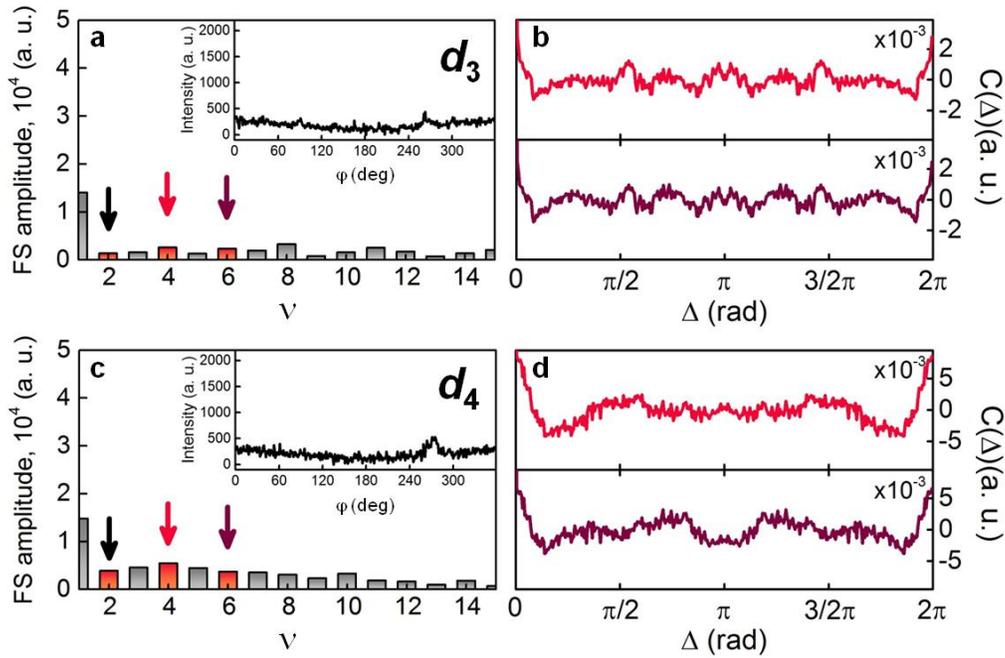

**Figure 4.** Fourier spectra and CCFs at $s_3$ and $s_4$ in the empty copper TEM grid. (a, c) Fourier spectra obtained by the Fourier transform of the angular intensity profile (insets, black curves). (b, d) Upper, red curves: CCF obtained by setting to zero the frequency ν=2 in the Fourier spectra. Lower, purple curves: CCF obtained by setting to zero the frequencies ν=2 and ν=4 in the Fourier spectra. Also in this case, no specific symmetries are retrieved, as opposed to the sample. The data at $s_1$ are not shown because the beam through the empty grid oversaturates the detector at very small angle, preventing the analysis of the data.

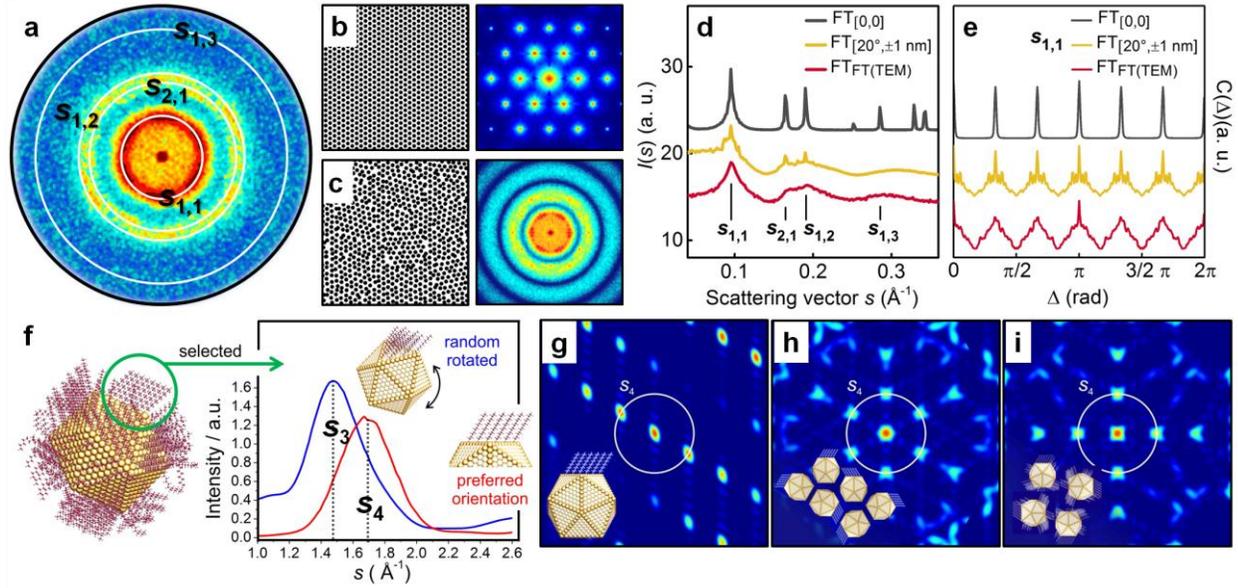

**Figure 5.** SLM and SASA simulations. (a) Squared amplitude of the Fourier transform of the experimental TEM image of the sample. Three different diffraction orders $n=1, 2, 3$ are detected from $s_1$ ($d_1$=6.6nm), together with the first order ($n=1$) from $s_2$ ($d_2$=3.8nm). The corresponding crystallographic planes in the supracrystal are marked in Fig. 1a, in the upper and lower insets, respectively. (b) NPs distribution simulated in SLM from a perfect lattice (domains rotation Gaussian distributed with mean=0°, standard deviation σ=0°, NPs displacement Δr=0 nm) and corresponding FT. (c) SLM simulation with σ=20°, Δr=±1nm and corresponding FT. (d) Radial averaged intensity $I(s)$ of the two SLM simulations. The grey profile is obtained from the FT of the SLM (σ=0°, Δr=0nm), and the yellow one from the SLM (σ=20°, Δr=±1nm). The $I(s)$ from the experimental FT of the TEM image is shown in purple. (e) CCFs at $s_{1,1}$ for the two SLM simulations compared to the FT of the experimental TEM image of the sample and color coded as in (d). (f-i) SASA simulation obtained using the following parameters: (i) Electron energy: 30keV, (ii) Sample-detector distance: 230mm, (iii) Sampling: 1000×1000 pixels, (iv) Pixel size in the detector plane: 50μm. (f) A single bunch of ordered ligands is selected for simulation. The radial intensity distribution shows a maximum at $s_3$=1.45Å$^{-1}$ for randomly oriented bunch of ligands and a maximum at $s_4$=1.68Å$^{-1}$ for oriented bunch of ligands. (g) The diffraction pattern simulated from single bunch of ligands of a NP in preferred orientation exhibits two peaks at $s_4$=1.68Å$^{-1}$. (h) Diffraction pattern simulated from ligands on the NPs surfaces that are in preferred orientation and arranged into a hexagonal lattice exhibits six peaks at $s_4$=1.68Å$^{-1}$. (i) Diffraction pattern simulated from ordered ligands on the NPs surfaces that are in preferred orientation and arranged into a tetragonal lattice exhibits four peaks at $s_4$=1.68Å$^{-1}$.

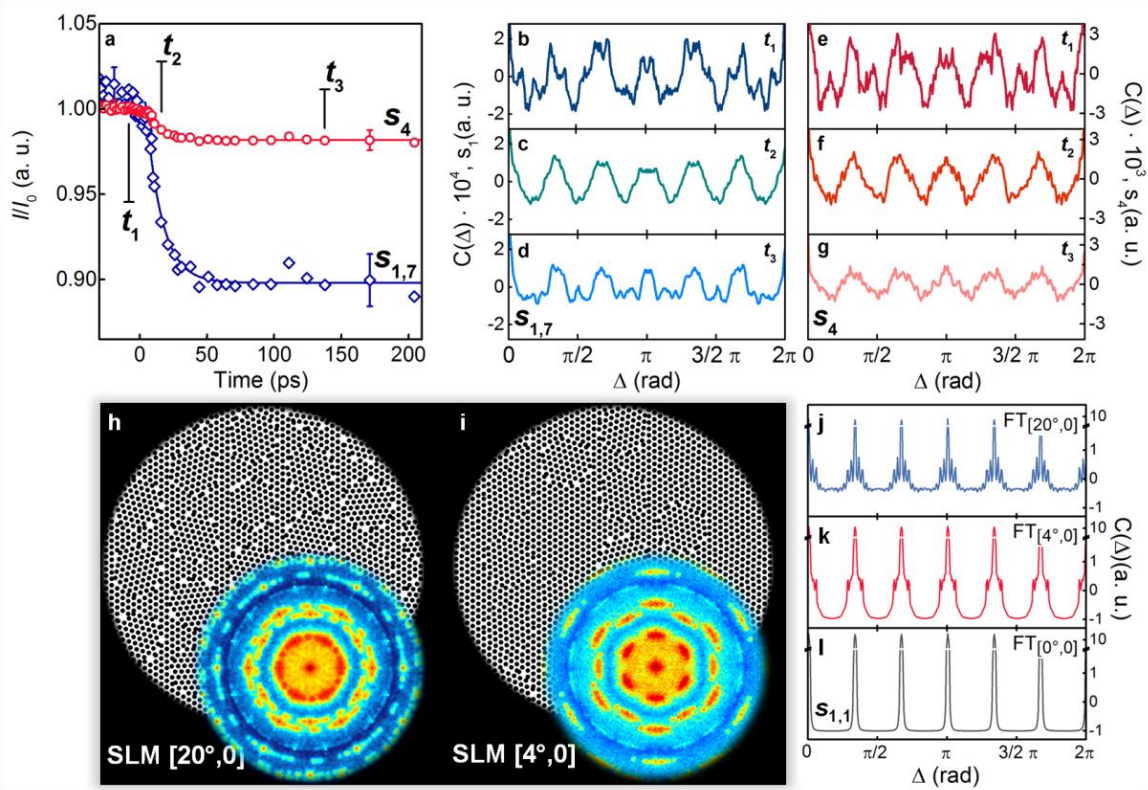

**Figure 6.** Small-angle ultrafast electron diffraction of dodecanethiol-capped gold NPs. (a) Dynamics of the radial intensity at $s_{1,7}$ (diamonds, blue trace) and $s_4$ (circles, red trace). Each intensity data set is normalized to its average value before time zero ($I_0$) and it has been fitted to a mono-exponential function. (b-g) CCFs at $s_{1,7}$ (b-d) and $s_4$ (e-g) at three time delays ($t_1$, $t_2$, $t_3$). The "preferential orientation" evidenced at negative times ($t_1$, panels b, e) is preserved upon photo-excitation. In fact, the six-fold symmetry in the CCF is clearly preserved shortly after ($t_2$, panels c, f) and more than 100ps after time-zero ($t_3$, panels d, g), despite the photo-induced disorder, which is reflected in a decrease in the CCF amplitude upon photo-excitation. (h) SLM simulation with σ=20°, Δr=0nm and corresponding FT. (i) SLM simulation with σ=4°, Δr=0nm and corresponding FT. (j-l) CCFs at $s_{1,1}$ for the SLM simulations [20°, 0] (blue), [4°, 0] (red) and for the perfect lattice model (grey) of Fig. 5b.

**Table 1.** The symmetries observed in the CCFs from the experimental diffraction pattern are assigned to the corresponding real-space objects. The modulations that are a signature of the NPs supracrystal are separated from the ones of the TEM grid supporting the supracrystal or the substrate.

|  | NP sample | | TEM grid + carbon | | empty copper TEM grid | |
|---|---|---|---|---|---|---|
|  | 4-fold | 6-fold | 4-fold | 6-fold | 4-fold | 6-fold |
| $S_1$ | tetragonal arrangement of the amorphous carbon in the substrate | the supracrystal structure | tetragonal arrangement of the amorphous carbon in the substrate | - | - | - |
| $S_3$ random orientation of NPs | tetragonal arrangement of the amorphous carbon in the substrate | - | tetragonal arrangement of the amorphous carbon in the substrate | - | - | - |
| $S_4$ preferred orientation of NPs | NPs local arrangement in the supracrystal | the supracrystal structure | - | - | - | - |

**References**

(1) Kim, Y.; Zhu, J.; Yeom, B.; Di Prima, M.; Su, X.; Kim, J.-G.; Yoo, S. J.; Uher, C.; Kotov, N. A. *Nature* **2013**, 500, 59–63.

(2) Pileni, M.-P. *Acc. Chem. Res.* **2007**, 40, 685–693.

(3) Talapin, D. V.; Lee, J.-S.; Kovalenko, M. V.; Shevchenko, E. V. *Chem. Rev.* **2010**, 110, 389–458.

(4) Feldheim, D. L.; Foss, C. A., Jr. *Metal nanoparticles, synthesis, characterization and applications* (CRC Press, 2001, New York, NY).

(5) Schmid, G. *Nanoparticles: from theory to application* (Wiley-VCH, 2006, Weinheim, Germany).

(6) Bishop, K. J. M.; Wilmer, C. E.; Soh, S.; Grzybowski, B. A. *Small* **2009**, 5, 1600–1630.

(7) Nakanishi, H.; Walker, D. A.; Bishop, K. J. M.; Wesson, P. J.; Yan, Y.; Soh, S.; Swaminathan, S.; Grzybowski, B. A. *Nat. Nanotechnol.* **2011**, 6, 740–746.

(8) De, M.; Ghosh, P. S.; Rotello, V. M. *Adv. Mater.* **2008**, 20, 4225–4241.

(9) Schäfer, S.; Liang, W.; Zewail, A. H. *Chem. Phys. Lett.* **2011**, 515, 278-282.

(10) Ernstorfer, R.; Harb, M.; Hebeisen, C. T.; Sciaini, G.; Dartigalongue, T.; Miller, R. J. D. *Science* **2009**, 323, 1033–1037.

(11) Chen, J.; Chen, W.-K.; Tang, J.; Rentzepis, P. M. *Proc. Natl. Acad. Sci. U. S. A.* **2011**, 108, 18887-18892.

(12) Chen, J.; Chen, W.-K.; Rentzepis, P. M. *J. Appl. Phys.* **2011**, 109, 113522.

(13) Ruan, C.-Y.; Murooka, Y.; Raman, R. K.; Murdick, R. A. *Nano Lett.* **2007**, 7, 1290-1296.

(14) Clark, J. N.; Beitra, L.; Xiong, G.; Higginbotham, A.; Fritz, D. M.; Lemke, H. T.; Zhu, D.; Chollet, M.; Williams, G. J.; Messerschmidt, M.; Abbey, B. R.; Harder, J.; Korsunsky, A. M.; Wark, J. S.; Robinson, I. K. *Science* **2013**, 341, 56-59.

(15) Carbone, F.; Musumeci, P.; Luiten, O. J.; Hebert, C. *Chem. Phys.* **2012**, 392, 1-9.

(16) Piazza, L.; Musumeci, P.; Luiten, O. J.; Carbone, F. *Micron* **2014**, 63, 40-46.

(17) Spence, J. C. H.; Weierstall, U.; Howells, M. *Ultramicroscopy* **2004**, 101, 149–152.

(18) Wochner, P.; Gutt, C.; Autenrieth, T.; Demmer, T.; Bugaev, V.; Díaz Ortiz, A.; Duri, A.; Zontone, F.; Grübel, G.; Dosch, H. *Proc. Natl. Acad. Sci. U. S. A.* **2009**, 106, 11511-11514.

(19) Wochner, P.; Castro-Colin, M.; Bogle, S. N.; Bugaev, V. N. *Int. J. Mat. Res.* **2011**, 102, 874-888.

(20) Latychevskaia, T.; Mancini, G. F.; Carbone, F. *Sci. Rep.* **2015**, 5, 16573.

(21) Mancini, G. F.; Mansart, B.; Pagano, S.; van der Geer, S.B.; de Loos, M.J.; Carbone, F. *Nucl. Instrum. Methods Phys. Res. A* **2012**, 691, 113-122.

(22) Ong, Q. K.; Reguera, J.; Silva, P. J.; Moglianetti, M.; Harkness, K.; Stellacci, F. *ACS Nano* **2013**, 7, 8529–8539.

(23) Zheng, N.; Fan, J; Stucky, G. D. *J. Am. Chem. Soc.* **2006**, 128, 6550-6551.

(24) Lee. Z.; Jeon, K.-J.; Dato, A.; Erni, R.; Richardson, T. J.; Frenklach, M.; Radmilovic V. *Nano Lett.* **2009**, 9, 3365–3369.

(25) Panthani, M. G.; Hessel, C. M.; Reid, D. K.; Asillas, G.; Jose-Yacaman, M.; Korgel, B. A. *J. Phys. Chem. C* **2012**, 116, 22463−22468.

(26) Gunawan, A. A.; Chernomordik, B. D.; Plemmons, D. S.; Deng, D. D.; Aydil, E. S.; Mkhoyan, K. A. *Chem. Mater.* **2014**, 26, 3328–3333.

(27) van Oudheusden, T.; Pasmans, P.L.E.M.; van der Geer, S.B.; de Loos, M.J.; van der Wiel, M.J.; Luiten, O.J. *Phys. Rev. Lett.* **2010**, 105, 264801.

(28) Kiewiet, F. B.; Kemper, A. H.; Luiten, O. J.; Brussaard, G. J. H.; van der Wiel, M. J. *Nucl. Instrum. Methods Phys. Res. A* **2002**, 484, 619-624.

(29) Musumeci, P.; Moody, J. T.; Scoby, C. M.; Gutierrez, M. S.; Westfall, M. *Appl. Phys. Lett.* **2010**, 97, 063502.

(30) Altarelli, M.; Kurta, R. P.; Vartanyants, I. A. *Phys. Rev. B* **2010**, 82, 104207.

(31) Kurta, R.; P. Altarelli, M.; Weckert, E.; Vartanyants I. A. *Phys. Rev. B* **2012**, 85, 184204.

(32) Badia, A.; Singh, S.; Demers, L.; Cuccia, L.; Brown, R.; Lennox, R. B. *Chem. Eur. J.* **1996**, 2, 359-363.

(33) Whetten, R. L.; Khoury, J. T.; Alvarez, M. M.; Murthy, S.; Vezmar, I.; Wang, Z. L.; Stephens, P. W.; Cleveland, C. L.; Luedtke, W. D.; Landman, U. *Adv. Mater.* **1996**, 8, 428-433.

(34) Badia, A.; Cuccia, L.; Demers, L.; Morin, F.; Lennox, R. B. *J. Am. Chem. Soc.* **1997**, 119, 2682-2692.

(35) NIST, electron elastic-scattering cross-section database 2000, NIST Standard Reference Database 71. *http://www.nist.gov/srd/nist64.cfm*

(36) Giamarchi, T. in *Encyclopedia of Complexity and Systems Science*, (Springer, 2009, New York, NY), pp. 2019-2038.

(37) Giamarchi, T.; Le Doussal, P. *Phys. Rev. B* **52**, 1242-1270 (1995).

(38) Klein, T.; Joumard, I.; Blanchard, S.; Marcus, J.; Cubitt, R.; Giamarchi, T.; Le Doussal, P. *Nature* **2001**, 413, 404-406.

# Supporting Information

**Order/Disorder Dynamics in a Dodecanethiol-Capped Gold Nanoparticles Supracrystal by Small-Angle Ultrafast Electron Diffraction**

**Authors:** Giulia Fulvia Mancini[a, 1], Tatiana Latychevskaia[b], Francesco Pennacchio[a], Javier Reguera[c, 2, 3], Francesco Stellacci[c], and Fabrizio Carbone[a, 4*]

**Affiliations**

[a]Laboratory for Ultrafast Microscopy and Electron Scattering, Lausanne Center for Ultrafast Science (LACUS), École Polytechnique Fédérale de Lausanne, CH-1015 Lausanne, Switzerland.

[b]Physics Institute, University of Zurich, Winterthurerstrasse 190, 8057 Zurich, Switzerland.

[c]Supramolecular Nanomaterials and Interfaces Laboratory, Institute of Materials, École Polytechnique Fédérale de Lausanne, CH-1015 Lausanne, Switzerland.

[1] Present address: JILA, University of Colorado, Boulder, CO 80309-0440.

[2] Present address: CIC biomaGUNE, Paseo de Miramón 182C, 20009 Donostia-San Sebastian, Spain.

[3] Present address: Ikerbasque, Basque Foundation for Science, 48011 Bilbao, Spain.

[4*]Prof. Fabrizio Carbone, Laboratory for Ultrafast Microscopy and Electron Scattering (LUMES). Faculty of Basic Sciences (SB), Institute for Condensed Matter Physics (ICMP), École Polytechnique Fédérale de Lausanne (EPFL), EPFL campus, CH H2 595, Station 6, CH-1015 Lausanne. Switzerland.

Email: fabrizio.carbone@epfl.ch

Phone: +41 21 69 30562

Spheres Lattice Model (SLM) simulations.

The SLM model is designed to match the arrangement of the NPs to that in TEM image of the sample. In both images, SLM and TEM, NPs are represented as opaque objects. Strictly speaking, NPs do not behave as opaque objects when interacting with electron waves. However, for the purpose of matching the NPs arrangement in the SLM and the TEM images by comparing their Fourier transforms, the phase distribution of the NPs can be neglected. Should one introduce phase-shifting properties of the NPs, it must be done in the same manner for both, SLM and TEM image, which will cause the same effect in their Fourier transforms. For smaller length scales, where the positions of individual atoms play a role, the SASA simulations are employed, which take into account three-dimensional positions of each atom and the related phase shifts.

Superposition of Atomic Scattering Amplitude (SASA) simulations and arrangement of ligands on the NP surface.

Gold NPs cores have a polyhedral morphology, where the truncated-octahedral shape is predominant together with icosahedra, dodecahedra and decahedra. Small displacements from the bulk positions occur in the surface atomic layers near edges and vertices, resulting in a slight rounding of the gold core shape[33]. The shape of the gold core for our model was selected to be icosahedral as this is the most common shape which gold nanocrystals made of a few thousands of atoms have at equilibrium[39, 40]. We remark that the most important parameter for the SASA simulation is the arrangement of the gold atoms on the facets of the nanoparticle, which is more or less the same for different geometries of the core, as that arrangement defines the distribution of the ligands and their atoms. The gold atoms are arranged in a face-centered cubic (*fcc*) lattice. In such a structure, the typical arrangements of atoms on the facets is the (111) surface, which is also the case for icosahedrally shaped nanostructures. For this reason the choice of a different model would not change the position of the maxima, or the diffraction peaks arrangement. Although the choice of an exact model for the shape of gold NPs is an interesting study by itself, which involves a whole independent branch of research[41, 42], it goes beyond the scope of the present study.

Dodecanethiol molecules bonded on the $Au_{(111)}$ facets of the NPs arrange preferably in the more thermodynamically stable *trans* isomers. *Trans* ligands among neighboring NPs interdigitate to compensate for the loss of density at the terminal part of the chains, while a minority of *gauche* chains occupy interstitial sites and domain boundaries[32, 34, 43, 44]. Therefore for the simulation we isolated a bunch of *trans* ligands considering the fact that, since only presence of an ordered structure can create a ring in diffraction pattern, the scattering from the ligands that are in *gauche* form, would only contribute to the background. Furthermore, the faceted shape of the NPs core leads to the interlocking of molecular bundles (*i. e.* groups of ligands on each facet) on neighboring particles.

**Table S1.** Diffraction orders from the crystallographic arrangement of NPs in the supracrystal. EDP is electron diffraction pattern, $FT_{TEM/SLM}$ is the Fourier transform of TEM and SLM model image respectively.

| Symbol | Scattering vector $s$ (Å$^{-1}$) | Real-space distances $d = 2\pi/s$ (nm) |
|---|---|---|
| $s_{1,1}$ (EDP, $FT_{TEM/SLM}$) | 0.09 | 6.6 |
| $s_{2,1}$ (EDP, $FT_{TEM/SLM}$) | 0.16 | 3.8 |
| $s_{1,2}$ (EDP, $FT_{TEM/SLM}$) | 0.19 | 6.6 |
| $s_{1,3}$ (EDP, $FT_{TEM/SLM}$) | 0.28 | 6.6 |
| $s_{1,7}$ (EDP) | 0.66 | 6.6 |

In the simulation, dodecanethiols were attached to the facets so that Au-S distance is 2.31Å[45]. The length of a dodecanethiol ligand chain completely stretched is 1.56nm. In the all-*trans* conformation the molecule is tilted 30° from the surface normal; it is also twisted with respect to the molecular backbone by 55° and far from the nearest-neighbor by 14°[46]. These ligands are known to form a superlattice commensurate with the Au-Au (111) distance of the underlying NP facets, which main distances are described in the main text.

Crystallographic arrangement of NPs in the supracrystal: diffraction and angular cross-correlation functions at $s_1$ for higher diffraction orders.

The diameter of the NPs core of 5.7nm as well as the NPs core-to-core distance $d_0$= 7.6nm has been determined by SLM simulations, as discussed in the main text. Given the hexagonal arrangement of the NPs in the supracrystal, and the results of the simulations, it follows that the crystallographic planes formed by the NPs in the supracrystal have a distance $d_1 = \frac{\sqrt{3}}{2} d_0 =$ 6.6nm and $d_2 = \frac{1}{2} d_0 =$ 3.8nm. The lowest order of diffraction comes from $d_1$. The crystallographic planes with distances $d_1$=6.6nm and $d_2$=3.8nm are depicted in the insets of Fig. 1a (see main text). The diffraction orders from the crystallographic arrangement of NPs in the supracrystal are summarized in Table S1. As already mentioned in the main text, all the diffraction orders of $s_1$ can be detected in the experiment, but the orders $s_{1,n}$ with $n<7$ are overexposed by the central beam. For this reason, the analysis of the diffraction from the crystallographic planes with $d_1$=6.6nm is carried out at the order $s_{1,7}$. The range of the Fourier transforms of both the TEM image and the SLM model image, is limited to $s<0.39$Å$^{-1}$ (see Fig. 5(a,c) in the main text). Thus, only the first three orders of $s_1$ ( $s_{1,n} = n \cdot \frac{2\pi}{d_1}$, with $n$ = 1, 2, 3), related to the crystallographic planes with distance $d_1$= 6.6nm are accessible.

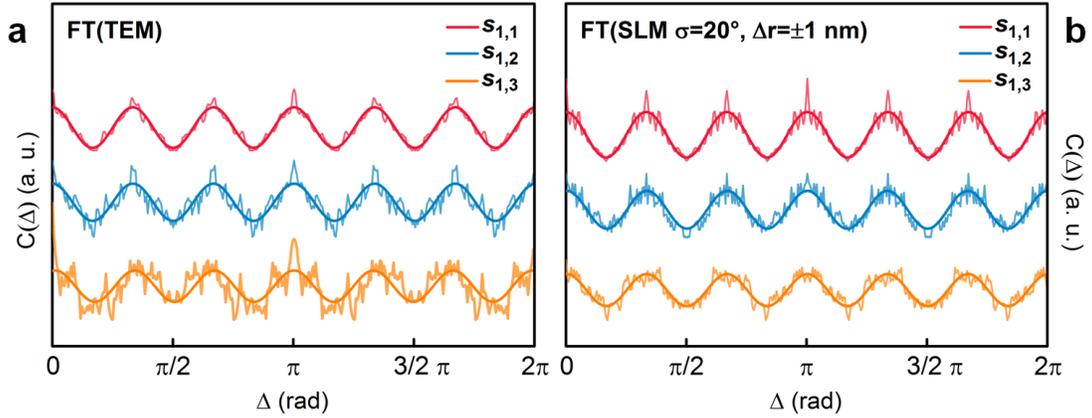

**Figure S1.** Comparison of the $C(\Delta)$ for the first three orders of $s_1$: $s_{1,1}$(red), $s_{1,2}$(blue), $s_{1,3}$(orange). (a) Results from the FT of the TEM image. (b) Results from the FT of the SLM ($\sigma=20°$, $\Delta r=\pm 1$ nm) simulation of the supracrystal.

These three diffraction orders have the same six-fold periodicity in the $C(\Delta)$, showing the same symmetry of the corresponding real-space object, as demonstrated by curves reported in Fig. S1a for the FT of the TEM image and in Fig. S1b for the FT of the SLM model with $\sigma = 20°$ and $\Delta r = \pm 1$nm.

Beam astigmatism and stability of the properties of the electron beam.

The strong two-fold character of the angular intensity curves at $s_1$, $s_3$ and $s_4$ is reflected by $\nu=2$ frequency in the Fourier spectrum. Figure S2 shows the Fourier spectra (panels a, c, e) and the CCF curves (panels b, d, f) at the three selected scattering vectors calculated from the raw intensity data spectrum, *i.e.* setting no frequency to 0. To separate the signal originating from the supracrystal from that of the support, we analyzed the data from an amorphous carbon-coated copper TEM grid (Fig. S3) and from an empty copper TEM grid (Fig. S4) following the same methodology. Both Figs. S3 and S4 show a predominant contribution of the low-order frequencies, $\nu=1$ and $\nu=2$, showing that this particular modulation is not specific to the sample and therefore, due to the slight astigmatism of the probing electron beam and to the presence of the amorphous carbon in the membrane supporting the NPs. The electron beam slight astigmatism is mostly due to stray magnetic fields which do not vary significantly between different experiments. In our experimental apparatus[21] the electron beam collimation and focusing solenoids are driven by high precision, high stability, current supplies in order to avoid magnetic fields fluctuations. The temporal chirp of the pulses is compensated by using a radiofrequency cavity oscillating on the $TM_{010}$ mode with a resonant frequency of 3 GHz, where the stability of the resonant frequency value sent to the cavity is controlled by a temperature control system with a precision of 4mK[27, 28].

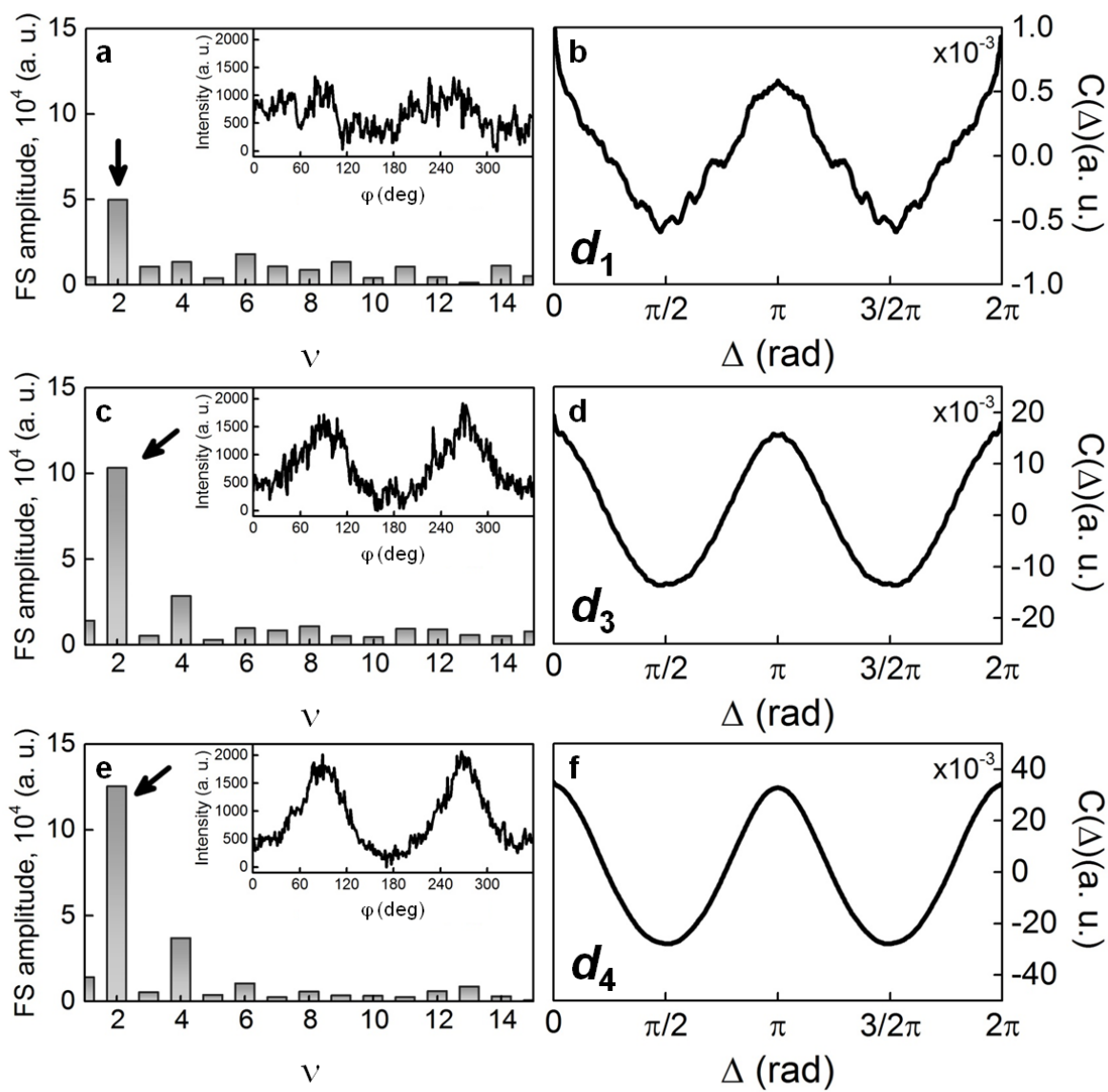

**Figure S2.** (a, c, e) Fourier spectra and (b, d, f) raw CCFs at $s_1$, $s_3$ and $s_4$ in the dodecanethiol-capped NPs supracrystal. The two-fold modulation is due to the beam astigmatism and the presence of amorphous carbon coating the sample substrate.

Once optimized the beam, characterized its stability and verified its astigmatism, the data sets from both the nanoparticles sample and the supporting substrate were taken in identical conditions in terms of electron flux, exposure time, experimental geometry and thus external magnetic fields.

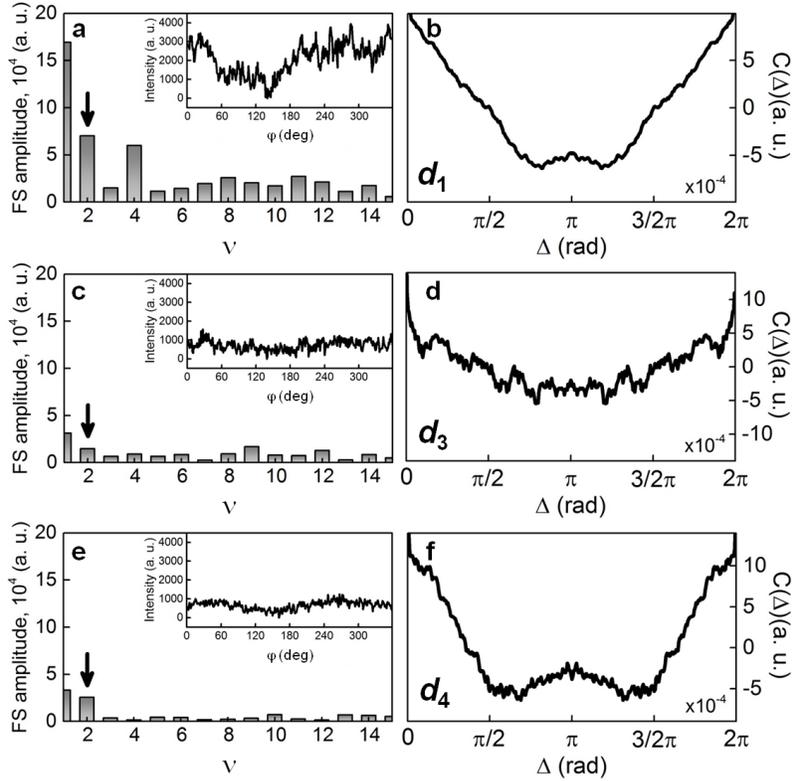

**Figure S3.** (a, c, e) Fourier spectra and (b, d, f) raw CCFs at $s_1$, $s_3$ and $s_4$ in the amorphous-carbon coated TEM grid. Effects of the beam astigmatism and the amorphous carbon are manifested in the low-frequency part of the spectrum.

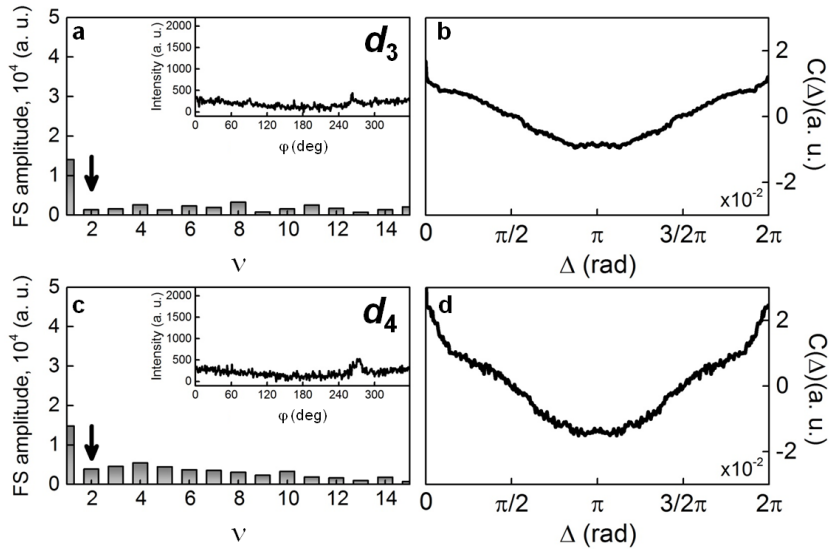

**Figure S4.** (a, c) Fourier spectra and (b, d) raw CCFs at $s_3$ and $s_4$ in the empty copper TEM grid. Effects of the beam astigmatism are manifested in the low-frequency part of the spectrum. The data at $s_1$ are not shown because the beam through the empty grid oversaturates the detector at very small angle, preventing the analysis of the data.

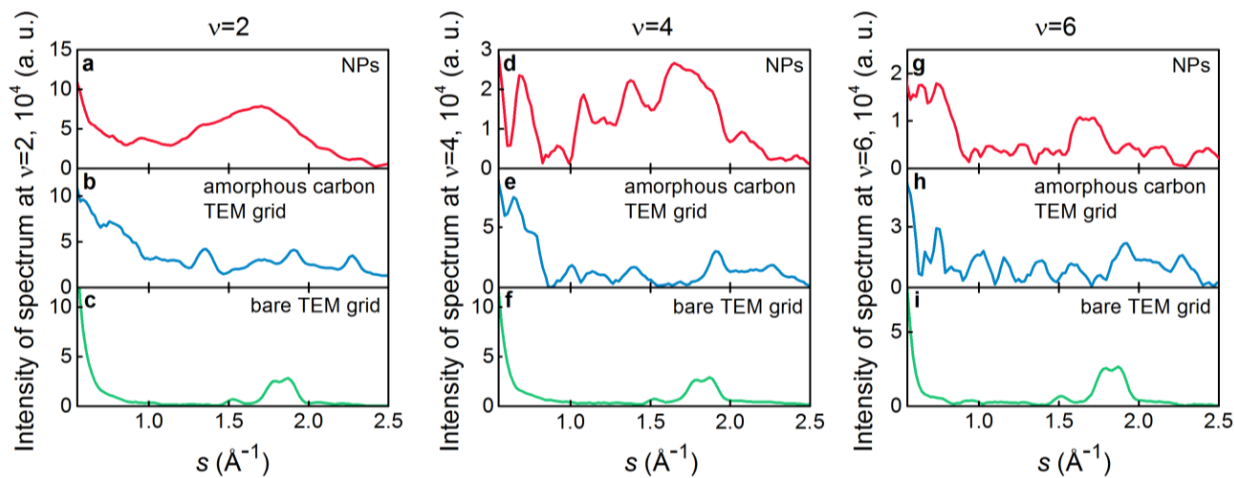

**Figure S5.** The amplitudes of the Fourier peaks at ν = 2, 4 and 6 as a function of *s* for three samples: (a, d, g) supracrystal of dodecanethiol-capped gold NPs deposited on the amorphous carbon-coated TEM grid (red traces), (b, e, h) amorphous carbon-coated TEM grid (blue traces) and (c, f, i) empty TEM copper grid (green traces).

Fourier components as function of *s*-vector coordinate.

In order to obtain the whole picture of the sample symmetries, and to separate them from the sample support ones, we plotted the amplitude of the Fourier components as a function of the scattering vector *s* for the three different samples: i) supracrystal of dodecanethiol-capped gold NPs deposited on the amorphous carbon-coated TEM grid, ii) amorphous carbon-coated TEM grid and iii) empty TEM copper grid. At each *s*-coordinate, the angular intensity was extracted and its Fourier spectrum was calculated. The amplitudes of the peaks at ν=2, 4 and 6 in the Fourier spectrum were considered. By repeating this procedure for all the *s*-coordinates the amplitudes of the Fourier peaks as a function of *s* were obtained, as shown in Fig. S5. From these plots we can draw the following conclusions:

(i) The empty TEM grid exhibits almost no distinct Fourier peaks and therefore, it adds no peaks and no related symmetries to the sample.

(ii) The amorphous carbon-coated TEM grid exhibits peaks in the ν=4 and ν=6 curves which can be attributed to the local symmetries in the amorphous carbon. The peaks observed in the ν=4 curve are explained by the presence of diamond-like carbon atoms arrangement. The peaks observed in the ν=6 curve are explained by the presence of graphene-like carbon atoms arrangement.

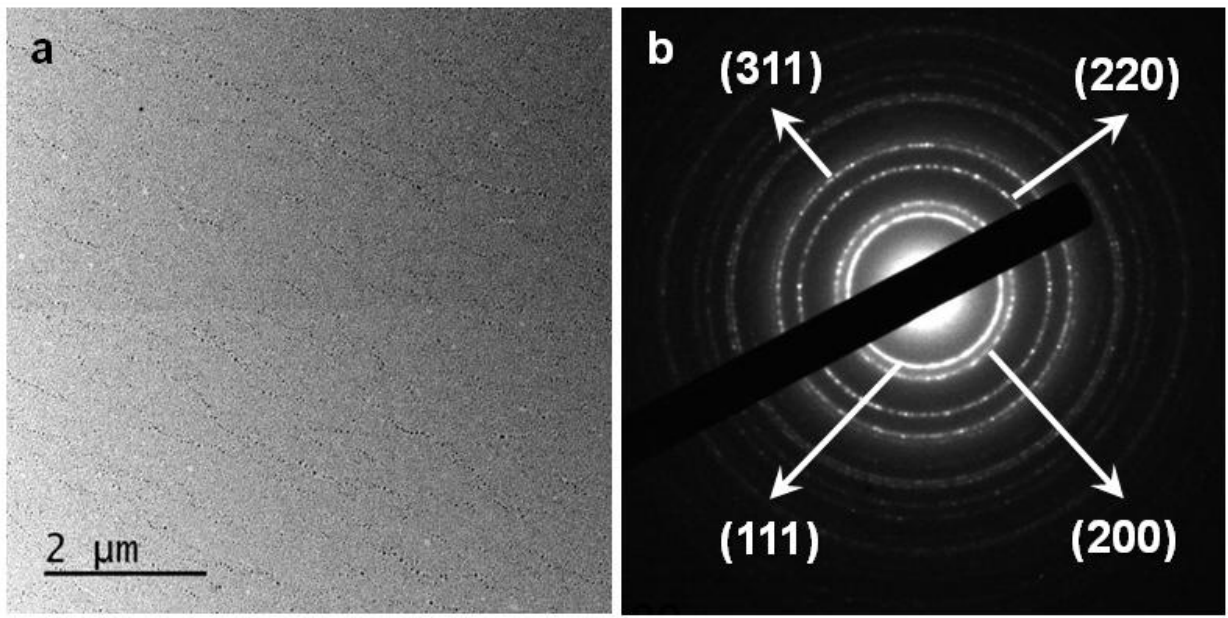

**Figure S6.** (a) TEM image of the supracrystal and (b) Polycrystalline diffraction pattern from the gold cores after the small-angle electron diffraction experiments.

(iii) Most of the peaks observed in the "amorphous carbon-coated TEM grid" curves and in the NPs curves do not share the same *s*-position, which allows separation between the signal from amorphous carbon and NPs.

(iv) The peak in the ν=6 dependency at $s=1.7\text{Å}^{-1}$ appears only for the NPs sample and thus can truly be related only to the NPs supracrystal. In the main text, this peak is attributed to the presence of "preferred orientations" of NPs. Also, in the ν=6 dependency for the NPs sample, the peak at $s=1.45\text{Å}^{-1}$ is clearly absent, which is an indication that the NPs are not randomly distributed.

Cross- validation of the absence of sample damaging.

After the pump-probe experiments on the dodecanethiol-capped gold NPs supracrystal, we carried out further TEM imaging and diffraction experiments to verify that the sample had not undergone damaging. Figure S6a shows a large field of view TEM image of the supracrystal, demonstrating that the NPs distribution and the overall integrity of the supracrystal have been preserved. The typical polycrystalline diffraction pattern from the gold cores is reported in Figure S6b. As already mentioned in the main text, our 200keV TEM is blind to the ligands. However, the use of a very low current and of a low repetition-rate allows the suprcrystal to fully recover between sequential shots.

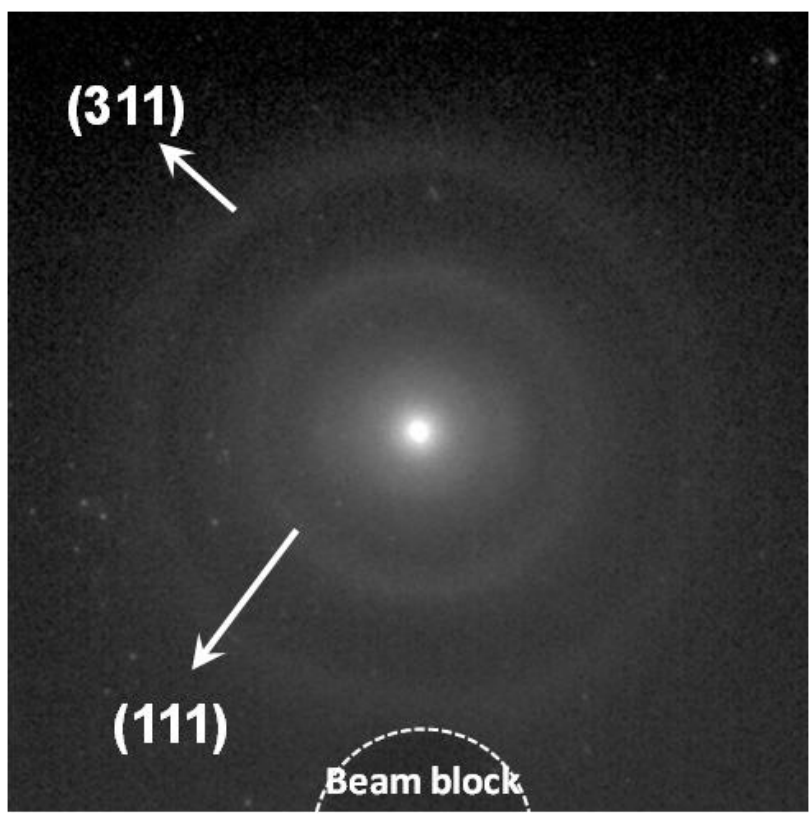

**Figure S7.** Raw small-angle electron diffraction pattern (logarithmic scale). This diffraction pattern is the result of the summation of 740 images acquired with the direct beam (without beam block), at very low counts ($2·10^5$ electrons for each image), to avoid saturation on the detector. The Debye-Scherrer rings from polycrystalline gold, marked in the figure, correspond to $s_{(111)}= 2.67$Å$^{-1}$ and $s_{(311)} = 5.10$Å$^{-1}$. The rings originate from the random orientation of the NPs in the supracrystal probed by the electron beam of 160μm spot-size.

The overall amount of charge from our probing electron packets on the sample, in relation to the overall sample exposure, is far below the limit threshold of $10^7$ Gy for biological sample damaging[15]. If we take into account $10^5$ electrons per pulse and the charge of a single electron ($1.602·10^{-19}$ C), the average charge per pulse on the sample is $1.602·10^{-2}$ pC. The total charge over time on the sample for a complete experiment has been estimated to 320.4pC/s, taking into account the 20kHz repetition-rate of our laser. In a scan of 56 delays at 740 images/delay, 4.11μC is the estimated total charge on the NPs, with this amount being distributed over several electron pulses impinging on the sample. Low sample damaging is therefore achieved using electrons, owing to their high cross-section for interaction with matter that allows reducing the accumulation times for efficient signal acquisition. Finally, following the methodology proposed in (47), the lack of radiation damage has been verified by checking the intensity of the diffraction rings originating from the gold cores.

# References

(39) Wang, Y.; Teitel, S.; Dellago, C. *Chem. Phys. Lett.* **2004**, 394, 257–261.

(40) Wang, Y.; Teitel, S.; Dellago, C. *J. Chem. Phys.* **2005**, 122, 214722.

(41) Daniel, M.-C.; Astruc, D. *Chem. Rev.* **2004**, 104, 293-346.

(42) Damasceno, P. F.; Engel, M.; Glotzer, S. C. *Science* **2012**, 337, 453-457.

(43) Badia, A.; Lennox, R. B.; Reven, L. *Acc. Chem. Res.* **2000**, 33, 475-481.

(44) Terrill, R. H.; Postlethwaite, T. A.; Chen, C.-H.; Poon, C.-D.; Terzis, A.; Chen, A.; Hutchison, J. E.; Clark, M. R.; Wignall, G.; Londono, J. D.; Superfine, R.; Falvo, M.; Johnson Jr., C. S.; Samulski, E. T.; Murray, R. W. *J. Am. Chem. Soc.* **1995**, 117, 12537-12548.

(45) Zanchet, D.; Tolentino, H.; Martins Alves, M. C.; Alves, O. L.; Ugarte, D. *Chem. Phys. Lett.* **2000**, 323, 167–172.

(46) Vericat, C.; Vela, M. E.; Salvarezza, R. C. *Phys. Chem. Chem. Phys.* **2005**, 7, 3258–3268.

(47) Howells, M.R.; Beetz, T.; Chapman, H.N.; Cui, C.; Holton, J. M.; Jacobsen, C.J.; Kirz, J.; Lima, E.; Marchesini, S.; Miao, H.; Sayre, D.; Shapiro, D. A.; Spence, J.C.H.; Starodub, D. *J. Electron Spectrosc. Relat. Phenom.* **2009**, 170, 4-12.